\documentclass{aa}
\usepackage{rotating}
\usepackage{graphicx}
\usepackage{txfonts}
\begin{document}

\title{9\,Sgr: uncovering an O-type spectroscopic binary \newline with an 8.6 year period\thanks{Based on observations collected at the European Southern Observatory (La Silla, Chile and Cerro Paranal, Chile) and the San Pedro M\'artir observatory (Mexico).}}
\author{G.\ Rauw\inst{1}\fnmsep\thanks{Honorary Research Associate FRS-FNRS (Belgium)} \and H.\ Sana\inst{2} \and M.\ Spano\inst{3} \and E.\ Gosset\inst{1}\fnmsep\thanks{Senior Research Associate FRS-FNRS (Belgium)} \and L.\ Mahy\inst{1}  \and M.\ De Becker\inst{1} \and P.\ Eenens\inst{4}}
\offprints{G.\ Rauw}
\mail{rauw@astro.ulg.ac.be}
\institute{Groupe d'Astrophysique des Hautes Energies, Institut d'Astrophysique et de G\'eophysique, Universit\'e de Li\`ege, All\'ee du 6 Ao\^ut, B\^at B5c, 4000 Li\`ege, Belgium
\and Sterrenkundig Instituut `Anton Pannekoek', Universiteit van Amsterdam, Science Park 904, 1098 XH Amsterdam, The Netherlands
\and Observatoire de Gen\`eve, Universit\'e de Gen\`eve, 51 Chemin des Maillettes, 1290, Sauverny, Switzerland
\and Departamento de Astronomia, Universidad de Guanajuato, Apartado 144, 36000 Guanajuato, GTO, Mexico}
\date{}
\abstract{The O-type object 9\,Sgr is a well-known synchrotron radio emitter. This feature is usually attributed to colliding-wind binary systems, but 9\,Sgr was long considered a single star.}{We have conducted a long-term spectroscopic monitoring of this star to investigate its multiplicity and search for evidence for wind-wind interactions.}{Radial velocities are determined and analysed using various period search methods. Spectral disentangling is applied to separate the spectra of the components of the binary system.}{We derive the first ever orbital solution of 9\,Sgr. The system is found to consist of an O3.5\,V((f$^+$)) primary and an O5-5.5\,V((f)) secondary moving around each other on a highly eccentric ($e = 0.7$), 8.6 year orbit. The spectra reveal no variable emission lines that could be formed in the wind interaction zone in agreement with the expected properties of the interaction in such a wide system.}{Our results provide further support to the paradigm of synchrotron radio emission from early-type stars being a manifestation of interacting winds in a binary system.}
\authorrunning{Rauw et al.}
\titlerunning{The O-type binary 9\,Sgr}
\maketitle
\section{Introduction \label{intro}}
It has been known for a while that a subset of the early-type stars of spectral-type O or Wolf-Rayet display a non-thermal radio emission in addition to the thermal free-free emission produced in their stellar winds (Abbott et al.\ \cite{ABC}, for a recent review see Blomme \cite{Blomme}). Dedicated multi-wavelength investigations of these objects have shown that they are mostly binary systems, hence suggesting that the acceleration of the relativistic electrons responsible for the synchrotron radio emission is related to the existence of hydrodynamical shocks between the stellar winds of the components of a massive binary system (e.g.\ Dougherty \& Williams \cite{DW}, De Becker \cite{Michael}, Naz\'e et al.\ \cite{CygOB2n9}, Sana et al.\ \cite{HD93250}). 

One of the first O-type stars discovered to display such a synchrotron radio emission was 9\,Sgr (= HD\,164794, Abbott et al.\ \cite{ABC}). This star is located inside the very young open cluster NGC\,6530 and contributes significantly to the ionization of the Lagoon Nebula (M\,8). Until recently, 9\,Sgr was considered as presumably single. However, its X-ray properties as well as episodic variations of some optical line profiles over long time-scales (Rauw et al.\ \cite{XMM}) suggested that it could be a binary system. In 2004, we observed another episode of optical line profile variations that we then interpreted as the partial deblending of the spectra around quadrature phase in a highly eccentric binary system with a period of about 8 -- 9\,years (Rauw et al.\ \cite{jenam})\footnote{There exists some confusion in the literature about the orbital period of 9\,Sgr. Indeed, the statement by Hubrig et al.\ (\cite{Hubrig1,Hubrig2}) of a 2.4\,yr orbital period, misquoting respectively Naz\'e et al.\ (\cite{CygOB2n9}) and Naz\'e et al.\ (\cite{Ofp}), is definitely wrong. Actually, whilst the work of Naz\'e et al.\ (\cite{CygOB2n9}) refers to another non-thermal radio emitter, Cyg\,OB2 \#9, which has indeed an orbital period of 2.4\,yr, this result as well as the paper of Naz\'e et al.\ (\cite{Ofp}) are totally unrelated to 9\,Sgr.}. 

Synchrotron radio emission requires the presence of relativistic electrons and a magnetic field. Concerning the latter, Hubrig et al.\ (\cite{Hubrig0}) claimed the detection of a magnetic field of about 200\,G in 9\,Sgr, although this field was only detected above the 3-$\sigma$ confidence level in one out of three observations of this star and the other two observations yielded a longitudinal field of opposite sign (which could be due to a different rotational phase). As to the origin of the relativistic electrons, the present paper aims at establishing the binary properties of 9\,Sgr to pave the way for future modelling efforts (e.g.\ Blomme et al.\ \cite{Blomme10}). 

\begin{figure*}
\resizebox{16cm}{!}{\includegraphics{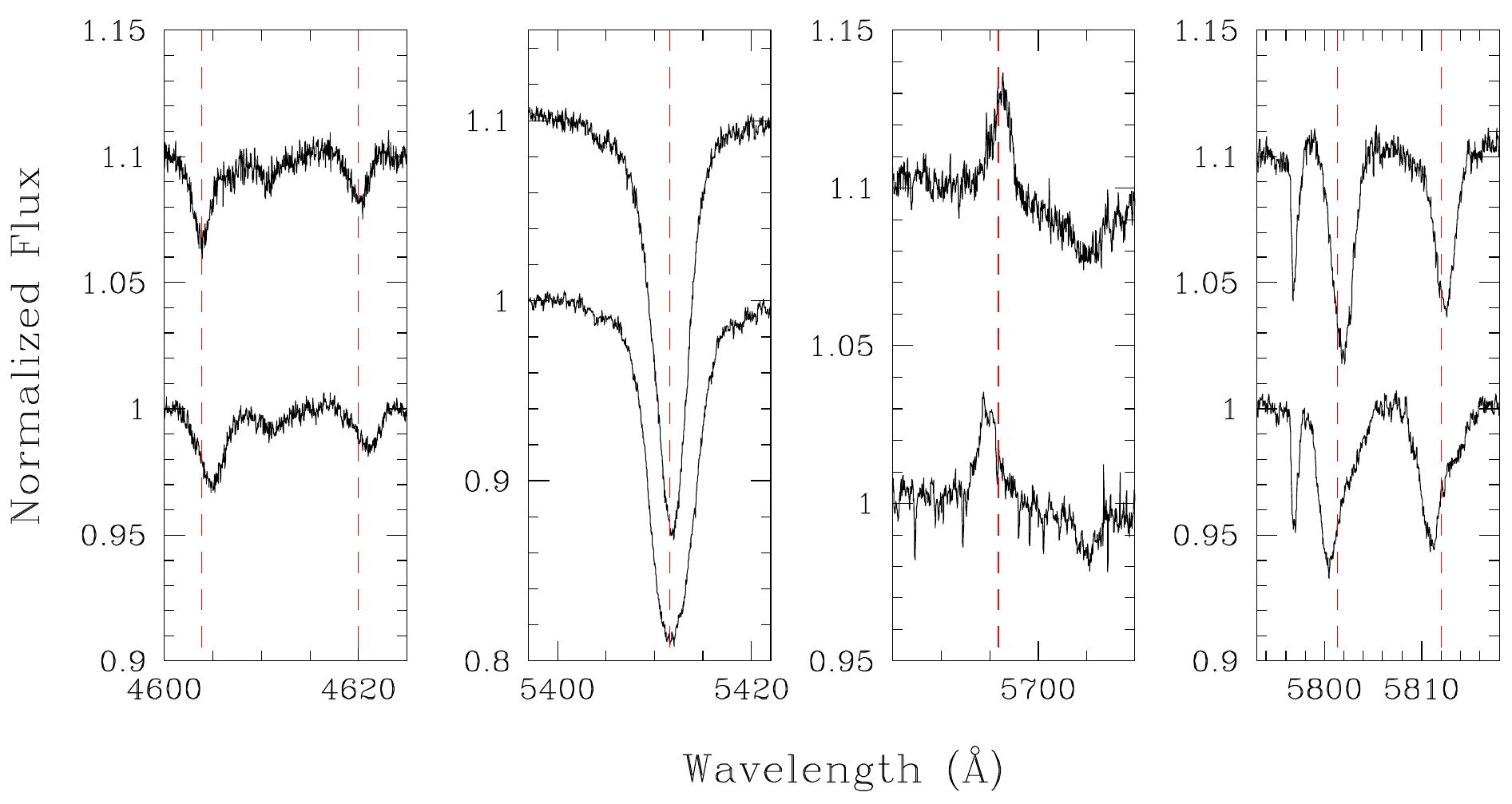}}
\caption{Various lines in the spectrum of 9\,Sgr observed at two different epochs. The lower spectrum was taken on 10 July 2004 (HJD\,2453196.586) whilst the uppermost spectrum was obtained on 5 October 2006 (HJD\,2454013.526). From left to right, these lines illustrate the various behaviours that exist in the spectrum: lines predominantly formed in the atmosphere of the primary (N\,{\sc v} $\lambda\lambda$\,4603, 4620), lines not showing any clear binary signature (He\,{\sc ii} $\lambda$\,5412), lines belonging to the secondary (C\,{\sc iii} $\lambda$\,5696) and lines displaying an SB2 signature (C\,{\sc iv} $\lambda\lambda$\,5801, 5812). In each panel, the dashed red lines indicate the rest wavelengths of the lines.\label{montage}}
\end{figure*}
\section{Observations}
Since our discovery of the probable binarity of 9\,Sgr, we monitored this star with a variety of telescopes and instruments. The dataset analyzed in this paper (see Table\,\ref{RVtab}) was taken over a timespan of twelve years.

The vast majority of our data was obtained at ESO at the La Silla observatory (Chile) with the FEROS \'echelle spectrograph on the ESO 1.5\,m telescope until October 2002 and on the MPG/ESO 2.2\,m telescope afterwards. FEROS (Kaufer et al.\ \cite{Kaufer}) has a spectral resolving power of 48000 and covers the spectral domain from slightly below 3800 to 9200\,\AA. The data were processed with an improved version of the FEROS pipeline working under MIDAS. The merged spectra were normalized over the spectral domain from 3900 to 6700\,\AA\ by fitting a spline function through a large number of carefully selected continuum windows. 

Fifteen additional spectra were obtained with the Coralie spectrograph (R = 55000) mounted on the Swiss 1.2\,m Leonhard Euler Telescope at La Silla. Coralie is an improved version of the Elodie spectrograph (Baranne et al.\ \cite{Baranne}) and covers the spectral range from 3850 to 6890\,\AA. The data were first reduced with the Coralie pipeline and subsequently normalized in the same manner as the FEROS data. 

One spectrum was obtained with the ESO-New Technology Telescope at La Silla equipped with the EMMI instrument in \'echelle mode (Dekker et al.\ \cite{DDD}). EMMI was operated with grating \#9 and grism \#3 as a cross-disperser to provide a resolving power of R = 7700 between about 4000 and 7200 \AA. The data were reduced with the \'echelle context of the MIDAS software and the normalization was done order by order using 3rd degree polynomials. 

Several observations of 9\,Sgr were taken with the UVES \'echelle spectrograph (D'Odorico et al.\ \cite{UVES}) on UT2 at the ESO Cerro Paranal observatory (Chile). UVES was used with the standard settings at 4370\,\AA\ (blue arm) and either of the standard settings at 5800, 6000 or 7600\,\AA\ (red arm). The slit width was set to 1\,arcsec, yielding a resolving power around 40000. The data were reduced with the UVES CPL pipeline and normalized in the same way as the FEROS spectra.

Another set of spectra was obtained with the Espresso spectrograph at the 2.12\,m telescope at the Observatorio Astron\'omico Nacional of San Pedro M\'artir (Mexico). The resolving power of this instrument is about 18000. The data reduction was done with the echelle context of the MIDAS software. Due to the strongly peaked blaze, the normalization was performed on a limited number of orders (around the most important spectral features) using spline functions adjusted to carefully chosen continuum windows.  

We finally used a set of archive spectra collected with the REOSC-SEL \'echelle spectrograph\footnote{On long-term loan from the Institute of Astrophysics, Li\`ege.} on the 2.15\,m Jorge Sahade telescope of Complejo Astron\'omico El Leoncito (Argentina). Again, normalization was found to be challenging due to the strongly peaked blaze function and was thus restricted to specific orders.

\section{Data analysis\label{general}}
Figure\,\ref{montage} illustrates some selected parts of the spectrum of 9\,Sgr at two different epochs. The changing position of the lines, and in some cases, the deblending of the lines are clearly seen. 
 
As a first step, we thus measured the radial velocities (hereafter RVs) of individual lines by fitting Gaussian profiles. For those lines that belong to only one component of the binary system (see below), a single Gaussian profile was fitted. For those lines that display a clear SB2 signature, we fitted two Gaussians whenever the lines were sufficiently deblended to provide a meaningful fit. This allowed us to obtain the RVs to be used in the orbital solution (see Sects.\,\ref{period} and \ref{orbit}) as well as in the disentangling process (see Sect.\,\ref{disent}).

We note a variety of behaviours (see Fig.\,\ref{montage}). Apart from some changes in the line width and depth, no clear signature of the orbital motion is found for the He\,{\sc ii} $\lambda\lambda$ 4200, 4542, 4686, 5412 lines. The He\,{\sc i} $\lambda\lambda$\,4471, 5876, O\,{\sc iii} $\lambda$\,5592, C\,{\sc iv} $\lambda\lambda$\,5801, 5812 absorption lines show a clear SB2 signature in our data taken in 2004 near maximum radial velocity separation, with the blue-shifted component being the stronger one (we shall refer to this star as the secondary in the following\footnote{This designation might seem strange at first sight, but it is in line with our orbital solution (see Sect.\,\ref{orbit}). In 9\,Sgr, the star with the strongest signature in He\,{\sc i} is the less massive and cooler component of the system.}). The same holds for a weak absorption at 4212\,\AA\ that we tentatively associate with Si\,{\sc iv}. On the other hand, the He\,{\sc i} $\lambda$\,4713 absorption and C\,{\sc iii} $\lambda$\,5696 emission move in phase with the secondary star but show no obvious SB2 signature. Finally, the Si\,{\sc iv} $\lambda$\,4116 emission, the N\,{\sc v} $\lambda\lambda$\,4603, 4620 absorptions as well as the N\,{\sc iv} $\lambda\lambda$\,5200, 5205 absorptions display an SB1 signature in phase with the primary. Therefore, we compute the RVs of the secondary as the mean of the measurements of the secondary components in He\,{\sc i} $\lambda\lambda$\,4471, 5876, O\,{\sc iii} $\lambda$\,5592, C\,{\sc iii}\,$\lambda$\,5696  and C\,{\sc iv}\,$\lambda\lambda$\,5801, 5812. All of these lines, except C\,{\sc iii}\,$\lambda$\,5696, display an SB2 signature near maximum separation, but the secondary's line always dominates the blend in such a way that we are rather confident that the mean value of these measurements should be a reliable indicator of the secondary's RV. In the same way, the primary RVs are obtained from the mean of the RVs of the Si\,{\sc iv} $\lambda$\,4116 and N\,{\sc v}\,$\lambda\lambda$\,4603, 4620 lines, which are probably free of blending with the secondary's lines. The results are listed in Table\,\ref{RVtab}. To quantify the uncertainty on the RV measurements, we have evaluated the 1-$\sigma$ dispersion about these means for each star, after correcting for systematic shifts between the radial velocities of some lines (see also Sect.\,\ref{orbit}).
 
The same partial line splitting as seen in our 2004 data (see Fig.\,\ref{montage}) was observed in 1987 by Fullerton (\cite{Alex}) and is also present in the REOSC spectra taken in 1995. This suggests an orbital period around 8.5\,years, making 9\,Sgr one of the spectroscopic O-type binaries with the longest orbital period so far. 

Williams et al.\ (\cite{Williams}) recently presented a set of 16 RV measurements of 9\,Sgr, obtained over 13 consecutive nights in May-June 2004, i.e.\ near maximum RV separation. These observations were taken in the blue (4058 -- 4732\,\AA) with the Ritchey-Chr\'etien spectrograph at the 1.5\,m telescope at CTIO with a resolving power of 2750. This set-up lacks the resolution that is needed to observe the SB2 signature even at maximum RV separation, explaining why Williams et al.\ (\cite{Williams}) did not observe the SB2 signature. 


\subsection{The orbital period \label{period}}
To constrain the orbital period of the system, we used the generalized Fourier periodogram technique proposed by Heck et al.\ (\cite{HMM}) and Gosset et al.\ (\cite{Gosset}). This method was applied to various combinations of the secondary RVs (our data, our data combined with the Fullerton \cite{Alex} data points) as well as the difference between the secondary and primary RVs. The best period is found to be $3165 \pm 110$\,days. The RV data folded with this period reveal a highly non-sinusoidal radial velocity curve indicating a rather large eccentricity. To check this result, we also applied the trial period method of Lafler \& Kinman (\cite{LK}). This method is a priori less sensitive to the shape of the RV curve. Again, using various combinations of the RVs, we obtain a best estimate of the period of $3181 \pm 110$\,days, which agrees quite well with the value obtained from the Fourier technique, given the uncertainty of the period\footnote{The uncertainty on the orbital frequency was estimated as one tenth of the natural width of the peaks in the Fourier periodogram as computed for the full set of secondary RVs.}. These estimates of the orbital period were subsequently used as input to the orbital solution code. We stress that the REOSC and Fullerton (\cite{Alex}) data actually have very limited weight in the period determination process. Indeed, not including these data changes the best estimate of the period by only about 15 days. However, including the old data points actually reduces the uncertainty on the period determination.

\subsection{Orbital solution \label{orbit}}
Using all the available RVs of the secondary component, we computed an SB1 orbital solution. For this purpose, we slightly shifted the RVs evaluated on the REOSC spectra as well as the data taken from Fullerton (\cite{Alex}) to account for a slight shift in the systemic velocities of the C\,{\sc iv} $\lambda\lambda$\,5801, 5812, He\,{\sc i} $\lambda$\,5876 lines used for the REOSC and Fullerton (\cite{Alex}) data on the one hand and the set of lines used to evaluate the secondary RVs from our new spectra on the other hand. Indeed, the six lines that we use in our determination of the secondary's RV yield systemic velocities that differ in the most extreme case by 16\,km\,s$^{-1}$ (C\,{\sc iii} $\lambda$\,5696 vs.\ C\,{\sc iv} $\lambda$\,5801). Since the REOSC and Fullerton (\cite{Alex}) RVs are drawn from a subset of the six lines, we have corrected these RVs by $+4.3$\,km\,s$^{-1}$, to avoid biasing our orbital solution.

The orbital solution was computed using the Li\`ege Orbital Solution Package (LOSP) code\footnote{http://staff.science.uva.nl/$\sim$hsana/losp.html} (Sana et al.\ \cite{SGR}) which is an improved version of the code originally proposed by Wolfe et al.\ (\cite{WHS}). 

\begin{figure}
\resizebox{9cm}{!}{\includegraphics{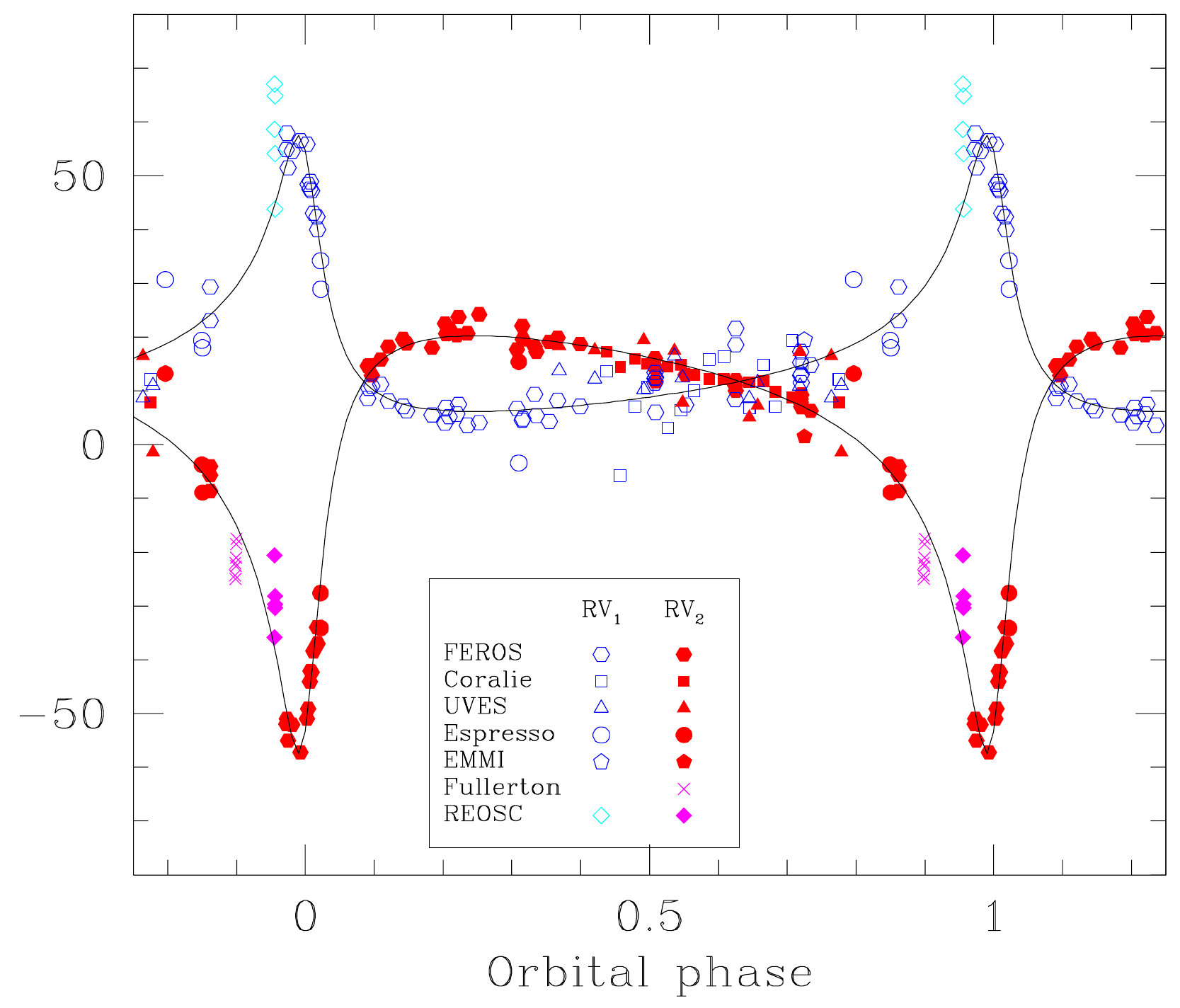}}
\resizebox{9cm}{!}{\includegraphics{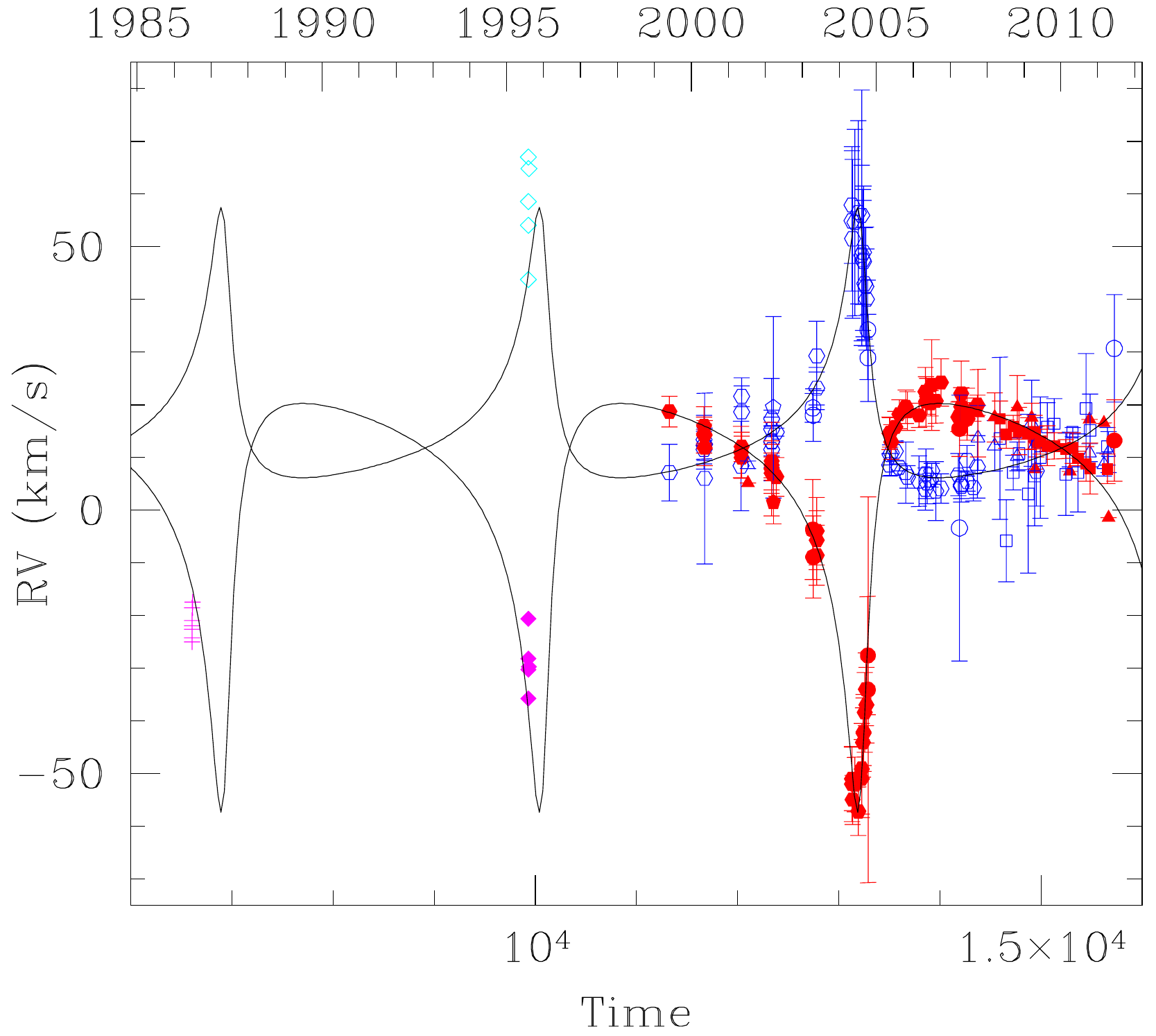}}
\caption{Radial velocity curve of 9\,Sgr. Open and filled symbols stand for the RVs of the primary and secondary star respectively. The different symbols indicate data taken with different instruments. The secondary RVs from Fullerton (\cite{Alex}) were not included in the computation of the orbital solution. The top panel yields the RV curve folded in orbital phase, whilst the lower panel illustrates the distribution as a function of time (given in years on the top axis and HJD - 2400000 on the bottom axis) of all RV measurements used in this paper.\label{solsb2}}
\end{figure}

For the SB1 orbital solution, the best-fit orbital period was found to be $3163\pm 12$\,days. This solution yields an eccentricity of $0.69$. We next used all 91 available measurements of the RVs of both stars (secondary and primary), to compute an SB2 orbital solution (see Fig.\,\ref{solsb2}). The best orbital period is now $3146 \pm 18$\,days and the eccentricity is essentially unchanged ($e = 0.70$). The parameters inferred from this orbital solution are listed in Table\,\ref{solorb}.  
 
\begin{table}[h!]
\caption{Orbital solution of 9\,Sgr \label{solorb}}
\begin{center}
\begin{tabular}{l c}
\hline
& All SB2 data  \\
\hline
Period (days)       & $3145.9 \pm 17.6$ \\
$T_0$ (HJD-2400000) & $46927 \pm 35$    \\
$e$                 & $0.695 \pm 0.011$ \\
$\omega$ ($^{\circ}$) & $26.4 \pm 1.3$   \\
$K_1$ (km\,s$^{-1}$) & $25.7 \pm 0.6$   \\
$K_2$ (km\,s$^{-1}$) & $38.8 \pm 0.9$   \\
$\gamma_1$ (km\,s$^{-1}$) & $15.8 \pm 0.4$ \\
$\gamma_2$ (km\,s$^{-1}$) &  $5.6 \pm 0.4$ \\
$a\,\sin{i}$ ($R_{\odot}$) & $2874 \pm 83$  \\
$q = m_2/m_1$           & $0.66 \pm 0.02$  \\
$m_1\,\sin^3{i}$ ($M_{\odot}$) & $19.5 \pm 1.4$ \\
$m_2\,\sin^3{i}$ ($M_{\odot}$) & $12.9 \pm 0.9$ \\
rms(O-C)$_1$ (km\,s$^{-1}$) & 4.2 \\
rms(O-C)$_2$ (km\,s$^{-1}$) & 3.2 \\
\hline
\end{tabular}
\tablefoot{$T_0$ stands for the time of periastron passage, $\omega$ is the primary star's longitude of periastron measured from the ascending node of the orbit. The quoted uncertainties correspond to 1-$\sigma$.}
\end{center}
\end{table}

We note the large difference in systemic velocity between the two stars. This is a common feature in massive binaries (e.g.\ Rauw et al.\ \cite{HDE}) and is usually interpreted as a signature of the stellar wind velocity fields affecting the positions of the lines. In the present case, this shift can also, at least partially, arise from the fact that we use different sets of lines to infer RV$_1$ and RV$_2$. Indeed, as pointed out above, there are systematic differences between the RVs of some of the lines. These differences remain roughly constant over the orbital cycle, except for the primary's Si\,{\sc iv} $\lambda$\,4116 emission and N\,{\sc v} $\lambda$\,4603 absorption lines near periastron passage. In fact, at these phases, the Si\,{\sc iv} line appears red-shifted by +14\,km\,s$^{-1}$ compared to RV$_1$ whilst the N\,{\sc v} line is simultaneously blue-shifted by $-11$\,km\,s$^{-1}$ again with respect to RV$_1$. This situation could indicate a perturbation of the primary's stellar wind due to enhanced wind-wind interactions when the stars are at periastron. The immediate consequence of this shift is the increase of the $\sigma_1$ dispersion around the phases of periastron passage (see Table\,\ref{RVtab}). We do not expect these effects to have a major impact on either the amplitudes of the RV curves or the results of the disentangling (see Sect.\,\ref{disent}). 

Based on our ephemerides, we predict that the next periastron passage of 9\,Sgr should occur in mid March 2013. 

\subsection{Disentangling \label{disent}}
We used our disentangling code based on the method of Gonz\'alez \& Levato (\cite{GL}) to separate the spectra of the secondary and primary component. This method iteratively reconstructs the secondary and primary spectra and allows in principle to simultaneously recover the RVs of both components. However, owing to its rather modest maximum RV separation, 9\,Sgr is definitely a very difficult case for any disentangling method. Indeed, in this system, the spectral lines are clearly double only at phases around the secondary's descending node (i.e.\ around periastron passage). The opposite configuration (secondary lines red-shifted) does not produce a sufficient RV excursion to separate the cores of the lines. This situation is expected to lead to an imperfect spectral disentangling, especially for the H\,{\sc i} and He\,{\sc ii} lines which have broad wings and do not show clear SB2 signatures around the secondary's descending node either (see Fig.\,\ref{montage}). Under these circumstances, it is not obvious that the disentangling method can actually improve the RV measurements. 

\begin{figure}[h]
\resizebox{8cm}{!}{\includegraphics{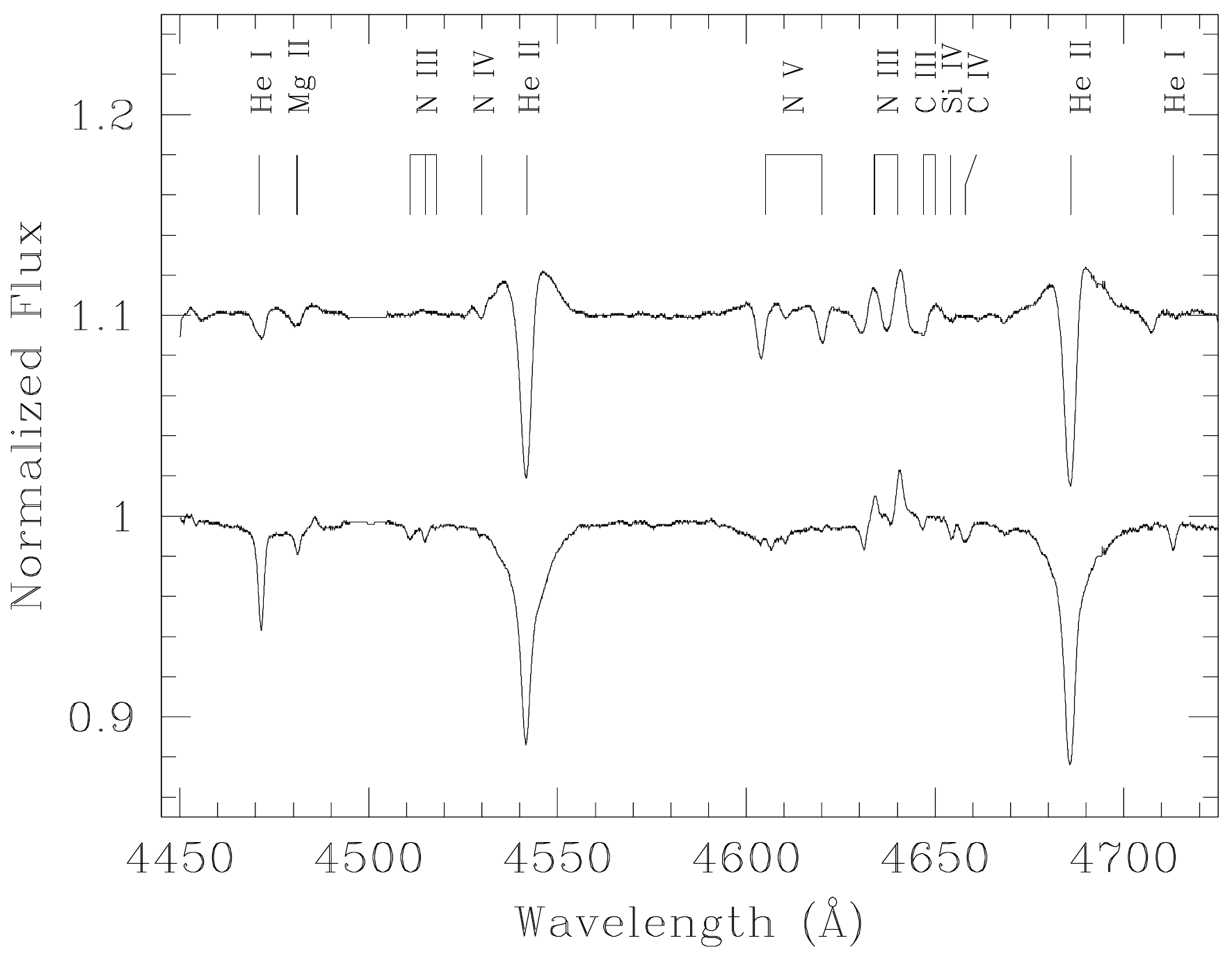}}
\resizebox{8cm}{!}{\includegraphics{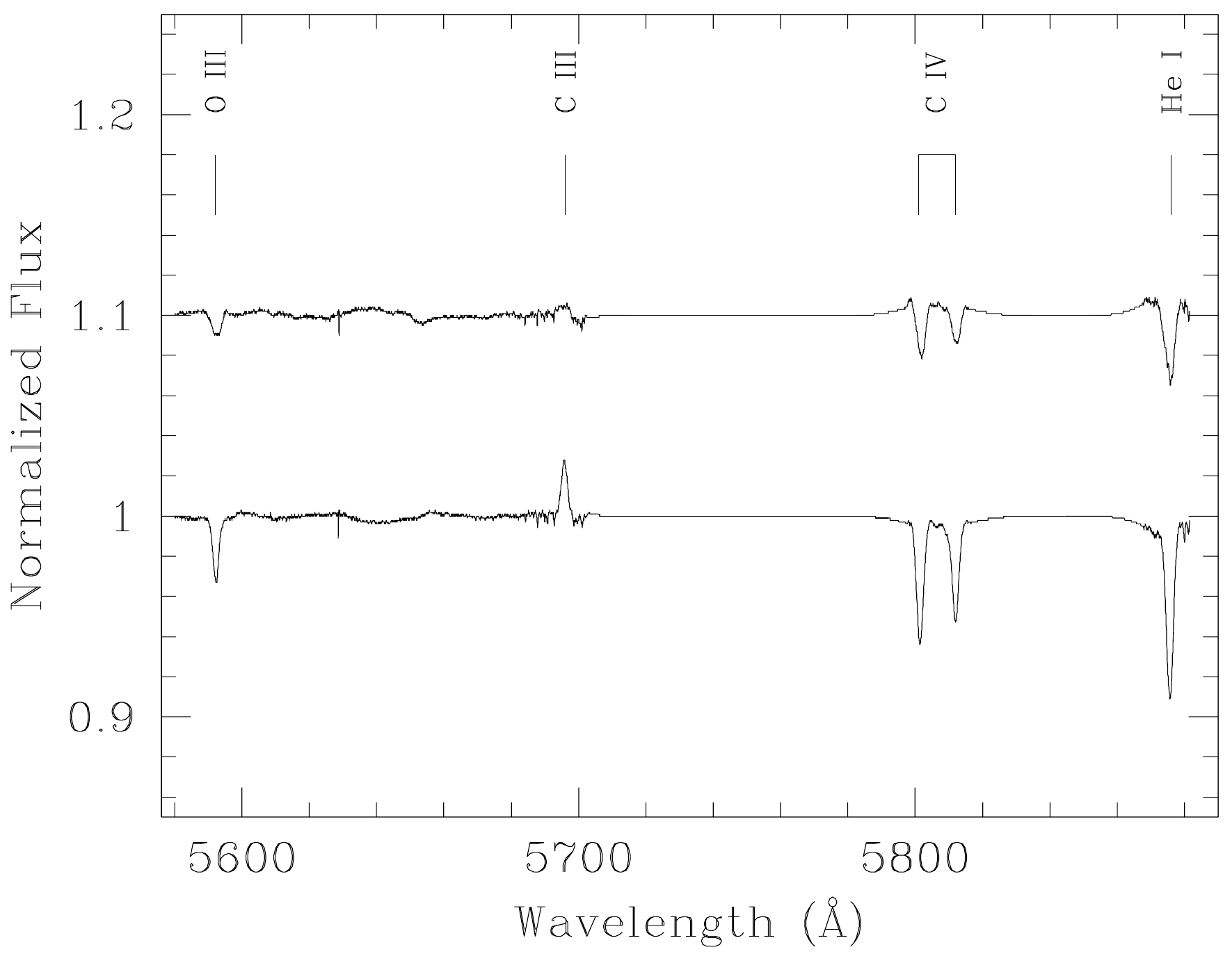}}
\caption{Disentangled spectra of the secondary and primary components of 9\,Sgr in the 4450 -- 4750 and 5580 -- 5890\,\AA\ regions. The normalization of the spectra is with respect to the level of the observed spectra and the primary spectrum is shifted vertically by 0.1 continuum units. Note that we excluded some parts of the yellow spectrum that are affected by strong diffuse interstellar bands. Note also that, because of the low RV amplitudes, there are a number of artefacts, such as the  wings of strong and broad lines that are not well represented.\label{disentspec}}
\end{figure}

To deal with as homogeneous a dataset as possible (in terms of spectral resolution and wavelength coverage) that covers most orbital phases, we applied the disentangling algorithm to our set of FEROS spectra only. We attempted disentangling with different treatments of the RVs, either keeping them fixed for one or both stars over part of the iterative process, or allowing the code to modify them. The result of our numerous tests is that, in the specific case of 9\,Sgr, disentangling does not improve the RVs and the best results on the spectral reconstruction are actually achieved by adopting the RVs determined from direct measurements of the lines (see above). We then applied the disentangling code on four wavelength domains: 4000 -- 4360, 4450 -- 4750, 5170 -- 5470, and 5580 -- 5890\AA. The results for two regions are illustrated in Fig.\,\ref{disentspec}. The disentangled spectra in Fig.\,\ref{disentspec} were not corrected for the optical brightness ratio and their normalization hence refers to the combined continua of both stars.

\subsection{Spectral classification and brightness ratio \label{classification}}
Unfortunately, the disentangling does not allow us to well recover the He\,{\sc ii} lines in the spectra of the components of 9\,Sgr. This is a serious handicap for spectral classification of O-type stars which relies to a large extent on the relative strength of He\,{\sc i} and He\,{\sc ii} lines. Nonetheless, Fig.\,\ref{disentspec} reveals a number of interesting features of the components of 9\,Sgr. For instance, we observe weak He\,{\sc i} absorptions in the spectrum of the primary, whilst these lines are much stronger in the spectrum of the secondary. Despite the uncertainties that affect the He\,{\sc ii} lines, we can obtain some useful information from these lines as well. First, He\,{\sc ii} $\lambda$\,4686 seems to be in relatively strong absorption in both stars, suggesting a main-sequence luminosity class for both of them. Second, He\,{\sc ii} $\lambda$\,4542 appears stronger than He\,{\sc i} $\lambda$\,4471 in both stars, indicating an early spectral type for both of them (Walborn \& Fitzpatrick \cite{WF}). 

The N\,{\sc iii} $\lambda\lambda$ 4634 -- 40 lines are apparently in emission in both stars, indicating that both should have an ((f)) tag. The absorption lines due to the highest ionization stages of nitrogen (N\,{\sc iv} and N\,{\sc v}) which are absent or very weak in the secondary's spectrum, are rather strong in the primary spectrum. Therefore, the primary is likely the intrinsically hotter component of the system. In this respect, we note also the presence of moderately strong Si\,{\sc iv} $\lambda\lambda$\,4088, 4116 emissions and of a very weak N\,{\sc iv} $\lambda$\,4058 emission in the spectrum of the primary. Overall, the primary spectrum resembles that of HD\,93128 classified as O3.5\,V((f$^+$)) in Walborn et al.\ (\cite{Walborn}) and we thus adopt the same classification for the primary star. The moderately strong C\,{\sc iii} $\lambda$\,5696 emission in the secondary spectrum and the rather strong O\,{\sc iii} and C\,{\sc iv} absorption lines are indicative of a spectral type later than O4 (Walborn \cite{Wal80}). Comparing the overall aspect of the secondary spectrum with the catalogue of Walborn et al.\ (\cite{Walborn}), we find that the best agreement is found with the spectrum of HD\,46150, an O5\,V((f)) star according to Walborn et al.\ (\cite{Walborn}) reclassified to O5.5\,V((f)) by Mahy et al.\ (\cite{Mahy}), and we thus adopt an O5-5.5\,V((f)) spectral classification for the secondary. Note that, according to the spectroscopic masses quoted by Martins et al.\ (\cite{Martins}), the dynamical mass ratio of $q = 0.66 \pm 0.02$ inferred from our orbital solution, is in reasonable agreement with the expected mass ratio ($0.66 - 0.73$) between an O5-5.5\,V secondary and an O3.5\,V primary. Also, comparing our minimum masses from Table\,\ref{solorb} with the spectroscopic masses quoted by Martins et al.\ (\cite{Martins}), we estimate an orbital inclination of $45^{\circ} \pm 1^{\circ}$. Given the rather wide separation between the two components of the binary, interferometry can resolve the system and should provide a determination of the separation of the stars projected on the sky. Indeed, assuming $i = 45^{\circ}$ and adopting a distance of 1.79\,kpc (see below), we predict angular separations of 3\,mas at periastron and 17\,mas at apastron. These numbers are within reach of modern interferometric facilities such as the VLTI. Combined with our orbital solution and an estimate of the distance, such a measurement yields, in principle, an independent determination of the orbital inclination and hence the absolute masses of the binary components.

From a preliminary attempt to disentangle the secondary and primary spectra, Rauw et al.\ (\cite{jenam}) inferred spectral types O4 for the primary and O7-8 for the secondary. Our new results are quite different. The reason for this discrepancy stems from the fact that our previous analysis was restricted to the region of the He\,{\sc i} $\lambda$\,5876 and C\,{\sc iv} $\lambda\lambda$\,5801, 5812 lines and we erroneously assumed the star with the strongest signature in these lines to be the hotter component of the system. As we have seen above, this is not the case.

We measured the equivalent widths (EWs) of some He\,{\sc i} and some metallic lines on the disentangled spectra of the two components (but referring to the combined continuum of the two stars). Comparing these numbers with the EWs observed in typical O5 and O5.5 stars as tabulated by Conti (\cite{Conti}) and Conti \& Frost (\cite{CF}), we estimated the dilution of the secondary's spectral signature by the light of the primary. Assuming an O5 spectral type for the secondary, we find that this star contributes $0.50 \pm 0.05$ of the total light in the optical domain. If we assume an O5.5 type instead, the contribution would be $0.39 \pm 0.04$. In other words, the secondary star should roughly contribute between 40 and 50\% of the total light. This would then imply an optical brightness ratio of about 1.0 -- 1.5 for the primary compared to the secondary\footnote{In our preliminary analysis (Rauw et al.\ \cite{jenam}), we inferred a brightness ratio of $4 \pm 1$ based on the equivalent width ratio of the He\,{\sc i} $\lambda$\,5876 and C\,{\sc iv} $\lambda\lambda$\,5801, 5812 lines. The discrepancy with our current result is entirely due to the different spectral types assumed in our previous work (see above).}. This result is in reasonable agreement with the expected difference in $M_V$ between an O3.5\,V and an O5-5.5\,V star. Indeed, from Table\,4 in Martins et al.\ (\cite{Martins}), we expect an optical brightness ratio in the range 1.5 -- 1.7. 

Unfortunately, there are not enough EWs of the earliest O-type stars tabulated in the literature, so that we cannot perform an independent evaluation of the brightness ratio on the EWs of the primary star.

If we assume that both stars have typical absolute magnitudes for their spectral types, we estimate a combined $M_V$ of $-6.20$. With an apparent $V$ magnitude of 5.93 and $B - V = 0.0$, we then estimate a distance modulus of 11.26 ($d = 1.79$\,kpc), in extremely good agreement with the distance of the open cluster NGC\,6530 ($1.78 \pm 0.08$\,kpc) as inferred by Sung et al.\ (\cite{Sung}).

We used the disentangled spectra to estimate the projected rotational velocities $v_{\rm eq}\,\sin{i}$ of the stars. For this purpose, we applied the Fourier transform method (Gray \cite{Gray}, Sim\'on-D\'{\i}az \& Herrero \cite{Sergio}) to the O\,{\sc iii} $\lambda$\,5592, C\,{\sc iv} $\lambda\lambda$\,5801, 5812 lines of the disentangled spectra. Indeed, these lines are the ones that are best reconstructed by the disentangling method. In some cases, the Fourier transforms of the lines are not very well represented by the theoretical Fourier transform of a rotational broadening function. However, by concentrating on the position of the first `zero' of the Fourier transform, we estimate $v_{\rm eq}\,\sin{i} = (87 \pm 16)$\,km\,s$^{-1}$ for the primary and $(57 \pm 5)$\,km\,s$^{-1}$ for the secondary. The quoted errors correspond to the dispersions about the mean of the values determined for the three lines and does not account for systematic uncertainties due e.g.\ to the presence of macroturbulence that could impact the measurement of the rotational velocity.

As a consistency check of the above results, we have simulated the synthetic spectra of an O3.5\,V star and an O5\,V star. For this purpose we have used the CMFGEN non-LTE atmosphere model (Hillier \& Miller \cite{CMFGEN}) which accounts for the presence of a stellar wind and includes the effects of line blanketing. In this model atmosphere code, the equations of radiative transfer are treated in the comoving frame and the hydrodynamic structure of the atmosphere is built from the TLUSTY models of the OSTAR2002 grid of Lanz \& Hubeny (\cite{TLUSTY}). We caution that our models are by no means adjusted to our disentangled spectra. Indeed, as pointed out above, there are artefacts in some lines of the disentangled spectra and furthermore, in our specific case, we lack reliable information on the mass-loss rates of the stars. However, these models were computed adopting some mean parameters of O3.5 and O5 dwarfs taken from Martins et al.\ (\cite{Martins}). For the primary, we assume $T_{\rm eff} = 43850$\,K, $\log{g} = 3.92$ (in cgs units), $\dot{M} = 9 \times 10^{-7}$\,$M_{\odot}$\,yr$^{-1}$, $v_{\infty} = 3500$\,km\,s$^{-1}$, whilst the secondary parameters were taken as $T_{\rm eff} = 40850$\,K, $\log{g} = 3.92$, $\dot{M} = 5 \times 10^{-7}$\,$M_{\odot}$\,yr$^{-1}$, $v_{\infty} = 3100$\,km\,s$^{-1}$. Each spectrum was further broadened by the observationally determined $v\,\sin{i}$. These synthetic spectra allowed us to check the brightness ratio and the behaviour of the He\,{\sc ii} lines. Concerning the first point, we evaluated the dilution factors of the spectra of each component by comparing its observed EWs to the simulated ones. In this way, we infer an optical brightness ratio (primary/secondary) of $0.87 \pm 0.40$ in reasonable agreement with the values derived hereabove. As to the second issue, by combining the two synthetic spectra shifted by the RVs at maximum separation, we recover the observed behaviours described in Sect.\,\ref{general} and Fig.\,\ref{montage}: the He\,{\sc i}, O\,{\sc iii} and C\,{\sc iv} lines are indeed found to display an SB2 signature whilst the He\,{\sc ii} lines, which have roughly equal strength in both stars, do not display a clear binary signature. Finally, we note that, in agreement with our observations, C\,{\sc iii} $\lambda$\,5696 emission is significantly present only in the synthetic O5\,V spectrum, whilst the CMFGEN spectra do not reproduce the observed N\,{\sc v} absorptions in the primary spectrum. Overall, the agreement is quite reasonable and lends further support to our spectral classification and optical brightness ratio near unity.

\section{Discussion}
Wind-wind interactions in massive binaries can leave their signature over a wide range of wavelength (see e.g.\ Rauw \cite{Brno} for a review). Among the most common signatures of this phenomenon are optical emission line variations, (variable) excess X-ray emission and non-thermal radio emission.

In close O-type binaries, the He\,{\sc ii} $\lambda$\,4686 and H$\alpha$ emission lines are frequently seen to display phase-locked variations due to the wind-wind collision (e.g.\ Linder et al.\ \cite{Linder} and references therein). However, in 9\,Sgr, these lines remain in absorption all over the orbital cycle and display little profile variations that are most easily explained by a partial deblending of the lines near periastron. In the H$\alpha$ line, the only emission is a weak narrow nebular emission arising from the Lagoon Nebula in which 9\,Sgr is embedded. This lack of optical emission lines is most likely due to the fact that the wind interaction in 9\,Sgr remains in the adiabatic regime over the entire orbital cycle. Indeed, in this case, the post-shock gas would not undergo radiative cooling and would not produce a high density region of sufficiently low temperature to produce significant line emission. There are two reasons why the wind interaction in 9\,Sgr would not be in the radiative regime. First the components of 9\,Sgr are still on the main-sequence and therefore feature rather low mass-loss rates. Second the separation between the two stars is quite large, even at periastron, where $r\,\sin{i} = 877$\,$R_{\odot}$ = 4\,AU, $r$ being the instantaneous separation between the stars. This implies that the pre-shock wind density is likely very low, preventing radiative cooling from being effective.

The {\it XMM-Newton} observation discussed by Rauw et al.\ (\cite{XMM}) was obtained on March 8, 2001, i.e.\ at orbital phase 0.605. We re-analysed this dataset in the light of the results presented above. The EPIC spectra can be fitted with a two-temperature thermal plasma model featuring a soft component with $kT_1 = 0.28$\,keV and a hard component at $kT_2 = 2.5$\,keV. The best-fit model has an X-ray flux (corrected for the interstellar absorption) of $4.3 \times 10^{-12}$\,erg\,cm$^{-2}$\,s$^{-1}$ over the 0.5 -- 10\,keV spectral domain. This number translates into an X-ray luminosity of $1.65 \times 10^{33}$\,erg\,s$^{-1}$ assuming a distance of 1.79\,kpc. As pointed out by Rauw et al.\ (\cite{XMM}), this X-ray luminosity is not exceptionally large, and does not provide direct evidence for a contribution of the wind-wind interaction. However, this observation was taken at phase 0.605, i.e.\ near apastron. If we assume that the wind interaction region in 9\,Sgr is in the adiabatic regime (see above), then the X-ray emission due to the colliding winds should vary as $1/r$ (Stevens et al.\ \cite{SPB}). Such a variation is indeed observed in some wide O-type binaries, a spectacular example being the 2.35\,yr period system Cyg\,OB2 \#9 (Naz\'e et al.\ 2012, submitted). If the entire X-ray emission of 9\,Sgr is due to the colliding wind interaction, a $1/r$ scaling would imply that the X-ray flux of 9\,Sgr should increase by a factor 5.4 near periastron ($r = 0.305\,a$) compared to our {\it XMM-Newton} observation ($r = 1.64\,a$). Alternatively, if we assume that only the harder spectral component seen in our EPIC spectra arises from the wind-wind interaction and hence undergoes the $1/r$ variation, we would still expect an increase of the observed X-ray flux by 50\% near periastron. A dedicated monitoring of the X-ray emission of the system around its next periastron would be most helpful to clarify this issue. 

As to the presence of relativistic electrons accelerated in the wind-wind collision zone, we stress that a VLA radio observation taken simultaneously with our {\it XMM} observation revealed a negative spectral index typical of non-thermal radio emission (Rauw et al.\ \cite{XMM}). Therefore, the shock between the winds is capable of accelerating electrons to relativistic velocities, even at phases close to apastron when the separation is largest. 
On the basis of the flux density measurements and on the radio spectral index between 3.6 and 6\,cm determined by Rauw et al.\ (\cite{XMM}), we estimate that the synchrotron luminosity at the epoch of the corresponding radio observations is of the order of 10$^{29}$\,erg\,s$^{-1}$. On the other hand, the cumulated kinetic power of the two stellar winds calculated from the quantities given at the end of Sect.\,\ref{classification} should be of the order of $5 \times 10^{36}$\,erg\,s$^{-1}$. This means that, at the orbital phase of the radio observation, a fraction of the order of $2 \times 10^{-8}$ of the kinetic power was converted into synchrotron radio luminosity. Considering first that at most a few percent of the kinetic power are injected in the wind-wind interaction, and second that a similar fraction of the injected power is used for particle acceleration, a power of about $10^{33}$\,erg\,s$^{-1}$ is available for relativistic particles, among which some are likely involved in non-thermal radiation processes. The measured synchrotron luminosity corresponds to the fractional power initially injected into relativistic electrons that is not radiated through inverse Compton scattering (the dominant energy loss process for relativistic electrons in colliding-wind binaries; see Pittard \& Dougherty \cite{PD}, De Becker \cite{Michael}). Assuming that the electron-to-proton ratio among the population of relativistic particles is of the order of 0.01 (as in the case of Galactic cosmic rays), converting about one percent of the energy of the relativistic electrons into synchrotron radio emission would be sufficient to account for the observations. Such a ratio is in agreement with the expected magnetic to radiation field energy densities in typical colliding-wind binaries (Pittard \& Dougherty \cite{PD}). Therefore, within our current understanding of non-thermal processes in massive binaries, the above considerations show that the synchrotron radiation of 9\,Sgr could indeed be entirely provided via an acceleration process taking place in the wind-wind interaction region of this binary system. The current uncertainties on the details of some of the energy conversion processes involved in the acceleration processes prevent us from doing a more extensive discussion in the present study. 
\section{Conclusions}
We have clearly established the multiplicity of 9\,Sgr and derived its first ever orbital solution. With an orbital period of 8.6 years, 9\,Sgr is one of a few O + O spectroscopic binaries with orbital periods longer than one year. Other well established O-type spectroscopic binaries in this part of the parameter space are HD\,15558 ($P = 440$\,days, De Becker et al.\ \cite{HD15558}), Cyg\,OB2 \#9 ($P = $858\,days, Naz\'e et al.\ \cite{CygOB2n9}, 2012) and 15\,Mon ($P = 9247$\,days, Gies et al.\ \cite{15Mon}). It is very likely that a number of similar systems remain to be discovered in this observational no-man's land. As demonstrated by our present study, exploring this part of the parameter space requires high-resolution spectroscopic monitoring over long periods of time. 

Our results support the current paradigm that attributes the presence of synchrotron radio emission in early-type stars to a wind-wind collision in a binary system. It is striking though that O-type non-thermal radio emitters discovered to be binaries over the last decade span a wide range of parameters. Indeed, the list of such stars includes systems with surprisingly short orbital periods such as Cyg\,OB2 \#8a ($P = 21.9$\,days, De Becker et al.\ \cite{CygOB2n8}) as well as much wider systems, the most extreme one so far being 9\,Sgr. At the same time, the list of non-thermal radio O-type stars includes main-sequence stars as well as giants or supergiants, i.e.\ stars having very different mass-loss rates. A characterization of these systems, both in terms of orbital and stellar parameters, is mandatory to properly understand the production of non-thermal electrons in O-star binaries and quantify their contribution to the acceleration of cosmic particles.

\acknowledgement{We wish to thank Dr.\ Ya\"el Naz\'e for taking the EMMI spectrum and for stimulating discussions on 9\,Sgr, Drs.\ Julia Arias and Nidia Morrell for providing us with the REOSC data, and an anonymous referee for his/her remarks and suggestions. We are grateful to the ESO service-mode staff for their help in collecting the FEROS and UVES data. The Li\`ege team acknowledges support from the Fonds de la Recherche Scientifique (FRS/FNRS), through the XMM/INTEGRAL PRODEX contract as well as by the Communaut\'e Fran\c caise de Belgique - Action de recherche concert\'ee - Acad\'emie Wallonie - Europe. PE acknowledges support from CONACYT.}

\appendix
\section{Journal of observations and radial velocities}
\begin{table*}[h!]
\caption{Journal of our observations of 9\,Sgr \label{RVtab}}
\begin{center}
\begin{tabular}{l c c c c c c c}
\hline
HJD & Exp.\ time & Instrument & RV$_1$ & $\sigma_1$ & RV$_2$ & $\sigma_2$ \\
    & (s)        &            & (km\,s$^{-1}$) & (km\,s$^{-1}$) & (km\,s$^{-1}$) & (km\,s$^{-1}$)\\
\hline
2451327.93054 & 450 & F &    7.1  &  5.4 &  18.7 & 2.9 \\
2451669.71311 & 420 & F &   11.5  &  3.7 &  14.5 & 1.8 \\
2451670.69770 & 240 & F &   12.3  &  2.6 &  16.0 & 3.6 \\
2451671.71623 & 360 & F &   13.4  &  4.6 &  14.3 & 3.4 \\
2451672.68625 & 300 & F &   12.5  &  3.1 &  14.3 & 2.7 \\
2451673.68442 & 480 & F &    6.0  & 16.2 &  11.8 & 3.4 \\
2452037.72550 & 600 & F &    8.4  &  8.5 &  11.0 & 4.9 \\
2452039.92845 & 300 & F &   21.6  &  2.0 &  10.0 & 4.0 \\
2452040.68815 & 540 & F &   18.6  &  6.5 &  12.0 & 2.8 \\
2452102.71686 &  90 & U &    8.6  & 0.7 &    5.1 & -- \\
2452335.89786 & 300 & F &   15.3  &  9.7 &   7.9 & 2.2 \\
2452336.88868 & 600 & F &   12.9  &  4.8 &   7.8 & 8.9\\
2452337.89775 & 600 & F &   17.3  &  1.7 &   8.5 & 4.7 \\
2452338.90100 & 600 & F &   13.1  &  2.8 &   9.3 & 3.2 \\
2452339.91296 & 600 & F &   11.5  &  4.5 &   7.0 & 3.8 \\
2452353.88979 &  90 & Em &  19.5  & 17.2 &   1.5 & 4.1 \\
2452381.87295 & 1200 & F &  14.8  &  0.8 &   6.3 & 3.7 \\
2452745.86397 &  450 & Es &  19.4  & 2.7 &  $-3.7$ &  9.5 \\
2452748.88352 &  540 & Es &  18.0  & 5.0 &  $-8.9$ &  7.8 \\
2452782.76326 & 700 & F &   23.1  &  3.1 & $-8.6$ & 5.7 \\
2452783.92065 & 900 & F &   23.1  &  4.0 & $-4.0$ & 6.4\\
2452784.93547 & 600 & F &   29.3  &  6.6 & $-5.7$ & 5.6\\
2453130.93332 & 240 & F &   54.9  & 13.3 & $-52.0$ & 7.1\\
2453133.88227 & 240 & F &   57.9  & 11.1 & $-51.0$ & 6.0\\
2453139.68916 & 450 & F &   51.5  & 15.1 & $-55.0$ & 4.7\\ 
2453159.73683 & 450 & F &   54.6  & 17.7 & $-52.1$ & 5.2\\
2453196.58561 & 450 & F &   56.5  & 17.4 & $-57.2$ & 4.6\\
2453226.64288 & 900 & F &   55.9  & 23.8 & $-51.0$ & 7.4\\
2453231.59984 & 450 & F &   48.4  & 17.1 & $-49.1$ & 8.6\\
2453240.59441 & 450 & F &   47.5  & 13.4 & $-44.1$ & 6.6\\
2453243.52397 & 450 & F &   48.9  &  9.9 & $-42.1$ & 6.4\\
2453246.58287 & 450 & F &   47.2  & 14.3 & $-42.3$ & 7.3\\   
2453255.52210 & 450 & F &   43.0  & 10.6 & $-38.4$ & 7.6\\
2453271.52285 & 300 & F &   42.4  & 11.3 & $-34.0$ & 6.9\\
2453273.51418 & 450 & F &   40.0  &  9.6 & $-37.0$ & 7.2\\
2453288.60031 &  840 & Es &  34.2  & 9.4 & $-27.6$ & 11.2 \\
2453289.59542 &  600 & Es &  28.9  & 8.3 & $-34.1$ & 36.6 \\
2453503.84288 & 450 & F &    8.6  & 2.1 &   14.6 & 3.1 \\
2453509.88418 & 300 & F &   10.5  & 1.9 &   14.7 & 2.8 \\
2453521.74907 & 450 & F &   10.9  & 4.3 &   12.9 & 3.3 \\
2453561.74156 & 450 & F &   11.2  & 1.9 &   15.8 & 3.0 \\
2453597.61875 & 450 & F &    8.1  & 0.7 &   18.2 & 2.7 \\
2453664.52625 & 450 & F &    7.1  & 5.8 &   19.6 & 3.2 \\
2453681.54040 & 450 & F &    6.3  & 2.1 &   18.8 & 3.7 \\
2453796.89444 & 300 & F &    5.5  & 4.8 &   18.0 & 2.3 \\
2453855.94364 & 300 & F &    4.0  & 4.0 &   22.5 & 4.6 \\
2453861.92973 & 300 & F &    6.9  & 4.7 &   20.6 & 4.7 \\
2453874.85744 & 600 & F &    5.2  & 4.0 &   21.6 & 3.4 \\
2453910.64138 & 800 & F &    5.7  & 1.3 &   20.3 & 2.4 \\
2453919.56510 & 600 & F &    7.5  & 1.4 &   23.7 & 8.6 \\
2453958.47874 & 600 & F &    3.6  & 5.6 &   20.7 & 4.2 \\
2454013.52591 & 414 & F &    4.1  & 2.8 &   24.2 & 4.5 \\
2454184.88276 & 552 & F &    6.7  & 4.0 &   17.7 & 2.8 \\
2454193.97917 &  900 & Es & $-3.4$ & 25.2 &   15.4  & 5.2 \\
2454210.93497 & 120 & F &    4.5  & 3.0 &   22.1 & 6.2 \\
2454212.82582 & 138 & F &    4.8  & 2.6 &   19.6 & 3.6 \\
2454267.72914 & 138 & F &    9.4  & 2.9 &   18.2 & 4.9 \\
2454274.86447 & 138 & F &    5.4  & 1.8 &   17.3 & 3.1 \\
\hline
\end{tabular}
\tablefoot{The codes for the instrument correspond to Coralie (C), FEROS (F), EMMI (Em), Espresso (Es) and UVES (U). The columns labelled $\sigma_1$ and $\sigma_2$ provide the 1-$\sigma$ dispersion about the mean RV of the lines that are used to evaluate the radial velocity of a given component (see Sect.\,\ref{general}). For a few UVES observations, no number is provided because the instrumental set-up covered only one of the lines used here.}
\end{center}
\end{table*}
\addtocounter{table}{-1}
\begin{table*}[h!]
\caption{continued}
\begin{center}
\begin{tabular}{l c c c c c c c}
\hline
HJD & Exp.\ time & Instrument & RV$_1$ & $\sigma_1$ & RV$_2$ & $\sigma_2$ \\
    & (s)        &            & (km\,s$^{-1}$) & (km\,s$^{-1}$) & (km\,s$^{-1}$) & (km\,s$^{-1}$)\\
\hline
2454333.53105 & 138 & F &    4.3  & 2.6 &   19.1 & 2.2 \\
2454372.53271 & 138 & F &    8.2  & 5.9 &   19.9 & 1.9 \\
2454378.49630 & 150 & U &   13.7  & 6.1 &   18.4 & 8.3 \\
2454540.82656 & 150 & U &   12.2  & 4.8 &   17.6 & 4.7 \\
2454595.83785 & 318 & C &   13.6  & 15.4 &  17.3 & 3.4 \\
2454658.67968 & 318 & C &  $-5.8$ &  7.8 &  14.4 & 2.7 \\
2454725.48261 & 565 & C &    7.1  &  3.3 &  15.9 & 2.1 \\
2454765.52334 & 150 & U &   10.2  & 2.7 &   19.5 & 6.0 \\ 
2454781.50546 & 565 & C &   10.7  &  3.0 &  15.0 & 4.5 \\
2454874.88694 & 600 & C &    3.1  & 15.1 &  14.5 & 3.0 \\
2454905.81592 & 150 & U &   16.5  & 8.2 &   17.5 & 2.8 \\
2454935.78120 & 624 & C &    6.4  &  9.4 &  14.8 & 3.2 \\
2454941.90945 &  30 & U &   12.4  & 5.1 &    7.8 & -- \\
2454956.93860 & 250 & F &    7.4  & 6.6 &   13.0 & 2.9 \\
2454995.62614 & 600 & C &   10.0  & 11.5 &  13.0 & 2.4 \\
2455064.56491 & 624 & C &   15.8  &  4.6 &  12.2 & 3.8 \\
2455132.49691 & 600 & C &   16.4  &  4.8 &  12.2 & 2.5 \\
2455247.89285 & 660 & C &    6.8  &  7.8 &  11.6 & 1.8 \\
2455284.89819 &  30 & U &   11.5  & 3.3 &    7.3 & -- \\  
2455311.89860 & 900 & C &   14.8  &  0.9 &  11.8 & 2.9 \\
2455365.75019 & 624 & C &    7.1  &  7.4 &   9.8 & 1.4 \\
2455443.58301 & 624 & C &   19.3  & 10.4 &   8.8 & 3.3 \\
2455478.53930 &  30 & U &   10.9  & 1.7 &   17.2 & -- \\
2455487.49425 & 600 & C &   15.1  &  6.9 &   7.9 & 4.8 \\
2455622.83675 &  30 & U &    8.7  & 2.2 &   16.5 & -- \\ 
2455656.88117 & 624 & C &   12.1  &  3.6 &   7.8 & 2.8 \\
2455668.90704 &  60 & U &   11.0  & 3.9 &  $-1.4$ & -- \\ 
2455724.71507 & 1200 & Es &  30.7  & 10.2 &   13.2  & 7.7 \\
\hline
\end{tabular}
\end{center}
\end{table*}

\end{document}